\documentclass[journal]{IEEEtai}

\usepackage[colorlinks,urlcolor=blue,linkcolor=blue,citecolor=blue]{hyperref}

\usepackage{color,array}
\usepackage{amssymb}
\usepackage{multirow}
\usepackage{graphicx}
\usepackage{subfigure}
\usepackage{tabularx}
\usepackage{algorithm}
\usepackage{array}
\usepackage{bm}
\usepackage{algorithmic}
\usepackage{subfigure}
\usepackage{hyperref}
\usepackage{amsmath}
\usepackage{booktabs}
\usepackage{multirow}
\usepackage{cite}
\usepackage{url}
\usepackage{color}
\usepackage[multiple]{footmisc}

\begin{document}

\title{HGFF: A Deep Reinforcement Learning Framework for Lifetime Maximization in Wireless Sensor Networks}  

\author{Xiaoxu Han, Xin Mu, and Jinghui Zhong, \IEEEmembership{Member, IEEE}
\thanks{Xiaoxu Han and Jinghui Zhong are with the South China University of Technology, Guangzhou 510006, China (e-mail: ftxiaoxu.han@mail.scut.edu.cn; jinghuizhong@scut.edu.cn). Xiaoxu Han is also with the Pengcheng Laboratory, Shenzhen 518000, China}

\thanks{Xin Mu is with the Pengcheng Laboratory, Shenzhen 518000, China (e-mail: mux@pcl.ac.cn).}

}

\maketitle

\begin{abstract}
Planning the movement of the sink to maximize the lifetime in wireless sensor networks is an essential problem of great research challenge and practical value. Many existing mobile sink techniques based on mathematical programming or heuristics have demonstrated the feasibility of the task. Nevertheless, the huge computation consumption or the over-reliance on human knowledge can result in relatively low performance.
In order to balance the need for high-quality solutions with the goal of minimizing inference time, we propose a new framework combining heterogeneous graph neural network with deep reinforcement learning to automatically construct the movement path of the sink. Modeling the wireless sensor networks as heterogeneous graphs, we utilize the graph neural network to learn representations of sites and sensors by aggregating features of neighbor nodes and extracting hierarchical graph features. 
Meanwhile, the multi-head attention mechanism is leveraged to allow the sites to attend to information from sensor nodes, which highly improves the expressive capacity of the learning model. Based on the node representations, a greedy policy is learned to append the next best site in the solution incrementally. We design ten types of static and dynamic maps to simulate different wireless sensor networks in the real world, and extensive experiments are conducted to evaluate and analyze our approach. The empirical results show that our approach consistently outperforms the existing methods on all types of maps. 

\end{abstract}

\begin{IEEEImpStatement}
Mobile sink methods are effective in prolonging the lifetime of wireless sensor networks (WSN). They balance the nodes' energy dissipation throughout the network. Planning the movement path of the sink is a critical challenge, as the states of each node in the network change dynamically. Previous methods based on mathematics or heuristics fail to find good solutions in a short time, due to the need for huge computation or suboptimal human knowledge. Our proposed approach overcomes these limitations. Firstly, features containing complex network topology and dynamic node information are extracted by the graph-based neural network. Secondly, the movement policy is learned automatically through reinforcement learning, without prior knowledge. Experiments were carried out under different network settings, and the results show that our method outperforms the previous methods significantly. The proposed method has notable contributions in maximizing the lifetime of WSN, thereby advancing its practical applications.
\end{IEEEImpStatement}

\begin{IEEEkeywords}
Wireless sensor network, Deep reinforcement learning, Heterogeneous graph neural network, Attention mechanism
\end{IEEEkeywords}

\section{Introduction}
\IEEEPARstart{O}ver the past few decades, Wireless Sensor Networks~(WSNs) have been widely applied in the real world, such as traffic control~\cite{nellore2016survey}, military target tracking~\cite{mahamuni2021intrusion}, and biomedical health monitoring~\cite{li2020computational}. 
A typical WSN consists of a sink and numerous sensor nodes that are used to measure information from the environment~(e.g. temperature, pressure, location, and so on). The data collected by sensor nodes are then transmitted to the sink through multi-hop communication, and eventually sent to the users from the sink. As the sensor nodes tend to be disposable and energy-constrained, the lifetime of WSN is limited and often defined as the duration time of the network until one or more sensor nodes run out of energy for the first time. Moreover, the inaccessibility of sensor deployment results in high recharging costs. Therefore, how to maximize the lifetime becomes one of the most important issues in the WSN community. 


Leveraging sink mobility has been shown to be an effective way to maximize the lifetime of WSNs~\cite{basagni2006new,agarwal2021survey}. 
Specifically, a mobile sink can move to various locations within the sensing region while collecting data from sensors. In this way, the sensors located near the sink change over time, which brings more balanced energy dissipation throughout the network. Thus, the lifetime of WSN with a mobile sink is prolonged. Efficiently scheduling the movement of the mobile sink is vital for the network's lifetime. 
Yet, determining the optimal traversal path for the sink remains a formidable challenge, known to be an NP-Hard problem~\cite{luo2006mobility}.


Various approaches for optimizing the sink's movement to prolong the WSN's lifetime have been proposed, including operational research-based~(OR) methods~\cite{yun2010maximizing,behdani2012decomposition,tashtarian2014maximizing} and heuristic methods~\cite{basagni2006new,liang2010prolonging,wang2018improved,maheshwari2021energy,zhong2012ant,zhong2019hyper}. 
Formulating the problem into a mathematical programming model~(e.g. the mixed integer linear programming model~(MILP)), the OR methods~(e.g.~branch-and-bound~(B\&B)) are usually guaranteed to find the optimal solution. 
Nevertheless, the exponential complexity of OR methods hinders their application to large-scale real-world instances. 
Heuristic methods can get satisfactory results within bearable time. 
Whereas well-designed and task-specific prior knowledge from human experts and trial-and-error are required, the optimality still cannot be guaranteed.

An alternative research direction for guiding the sink's path to maximize WSN's lifetime is based on Reinforcement Learning~(RL). RL methods have been shown to be able to automatically learn good heuristics
within brief solving time, as demonstrated by its outstanding performance in tackling combinational optimization problems, like Traveling Salesman Problem~(TSP)~\cite{DBLP:conf/iclr/BelloPL0B17}, Vehicle Routing Problem~(VRP)~\cite{DBLP:conf/iclr/KoolHW19} and so on~\cite{khalil2017learning,barrett2020exploratory,bengio2021machine,9471008}. 
Without prior knowledge and labeled data, RL can learn effective policy from scratch. 
Given that the movement of mobile sink can be modeled as a Markov decision process~(MDP), several works for planning the traverse path of sink based on RL have been proposed~\cite{forster2009clique,mustapha2017energy,soni2018novel,krishnan2021reinforcement}.
They leverage the tabular-based Q-learning method~\cite{watkins1992q} to find the optimal path, where the Q value is updated using the Bellman equation during trial-and-error. While the existing RL-based works still face the following technical challenges: 

$\bullet$ {\bf Curse of dimensionality and continuous spaces}. 
Maintaining Q values in a table for every possible state-action pair can become infeasible for large or continuous state spaces. Thus, they can not handle the high-dimensional space brought by large-scale WSNs, which may consist of hundreds or thousands of nodes. Besides, information loss is caused by the inability to handle continuous data input of WSNs.

$\bullet$ {\bf Unable to handle dynamic scenario}. 
In dynamic WSNs, the location of sensors may change at any time. While the existing methods can not adapt well to changes without further exploration. In addition, they lack the ability to generalization and have difficulty processing unseen dynamic scenarios.

To tackle the issues and limitations above, we propose a novel Deep Reinforcement Learning~(DRL) framework with heterogeneous graph-based feature fusion~(HGFF\footnote{Source codes are available in:~\url{https://github.com/xiaoxuh/HGFF_Maximize-the-lifetime-of-WSN-with-DRL}.}) to address the lifetime maximization problem in WSNs. We first abstract the problem as an optimization on the heterogeneous graph with two types of nodes, i.e. sites and sensors.
Then, extracting features from the original WSNs graph input, the graph neural network captures the information of the node in relation to its surrounding graph nodes.
Specifically, the node representations are learned using a combination of learnable type embedding and message-passing aggregation. The introduction of type embedding is to learn the heterogeneity of node features and better exploit graph structure based on the node type information. Meanwhile, an attention-based global feature fusion operation is proposed to enable the site nodes to attend over features of all the sensor nodes in the graph. We use Double Q-learning~\cite{hasselt2010double} method to learn greedy policy for guiding the movement of the sink. Following the algorithm pattern first produced in~\cite{khalil2017learning}, our framework constructs the route of mobile sink incrementally by greedily selecting the site node with the highest Q value, which is estimated based on the node embeddings. 
To demonstrate the superiority of the proposed method across different WSNs, we compare HGFF with different kinds of existing methods on a series of types of WSNs.  
The empirical results show that HGFF achieves significant improvement in prolonging the lifetime of WSNs.

Our main contributions in this research are as follows:
\begin{enumerate}
	\item We present a new deep reinforcement learning framework HGFF for lifetime maximization in WSN. HGFF generates effective heuristics to guide the movement of the sink in an end-to-end manner by DRL, which requires neither huge computation consumption nor human knowledge.
	
	\item We model a WSN as a heterogeneous graph, and use graph-based feature fusion techniques to enhance the representations for nodes. To be specific, learnable type embedding is incorporated into the graph neural network to learn the heterogeneous information of different types of nodes. Moreover, different from the neighbor node aggregation in common graph attention networks, we adopt a global-based attention mechanism to further extract the relativity information.  The improved node representation thus improving the quality of the solution.
 
	\item We conduct extensive experiments to verify the effectiveness of our proposed approach. The empirical results have shown that HGFF consistently outperforms other methods across various types of maps, including challenging
	large-scale networks and dynamic WSNs.

\end{enumerate}

The rest of the paper is structured as follows. In Section~\ref{section2}, we review different research using mathematical methods, heuristic methods, and RL-based methods in the problem of lifetime maximization WSN context. Then in Section~\ref{section3}, we define the mathematical model of the movement problem for the sink. In Section~\ref{section4}, we present the MDP formulation and introduce our framework in detail.  Section~\ref{section5} presents the experimental settings for evaluating our proposed method and the empirical results, which includes the setup of experiments, the results, and the analysis of our method with existing methods. Finally, we conclude the paper and look forward to future work in Section~\ref{section6}.

\section{Related Work}\label{section2}

Due to the limited energy of sensors, how to improve energy using efficiency is a critical issue in WSNs, which is directly related to the life span of the entire network. Static sink has been shown to perform badly in achieving energy efficiency in collecting data in wireless sensor networks~\cite{khan2013static}. The main reason for this phenomenon is the ``energy hole'' problem~\cite{ren2015lifetime}, namely, the sensor nodes around the sink are prematurely depleted of energy due to the need to forward more data, which is likely to form a sensor network energy consumption bottleneck, resulting in a short lifetime. Therefore, the mobile sink has been adopted by various methods to improve the lifetime of wireless sensor networks.

The optimization problem for scheduling the movement of the sink to maximize lifetime can be modeled as different mathematical optimization models. Leveraging the MILP model, Basagni~\emph{et al.}\cite{basagni2006new} attempt to maximize the lifetime of WSNs with constraints including a movement restriction for the sink traveling between two different sites, and the energy cost of the mobile sink in scheduling. Setting an objective to maximize the accumulated sojourn time of the mobile sink in event-driven applications, a convex optimization model to determine the optimal trajectory of a mobile sink without relying on predefined structures is introduced in~\cite{tashtarian2014maximizing}. In order to make the optimization model closer to reality, more elements are considered in the mathematical problem. The storage capacity of sensors is considered in~\cite{yun2010maximizing}, for the case of delayed data delivery is permitted. The authors propose a more complicated model based on Linear Programming~(LP), which decides the data delivery of sensors according to whether the sink is in the position most conducive to achieving the longest network lifetime. Behdani~\emph{et al.} \cite{behdani2012decomposition} propose a column generation algorithm for both prohibits delay tolerance and allows delay tolerance wireless sensor networks. Although the mathematical optimization models are bound to get a high-quality solution, the consuming time and computing power costs are unbearable for large-scale problems.

Some works concentrate on designing heuristic methods manually to solve the problem. A greedy maximum residual energy~(GMRE) heuristic rule is proposed in~\cite{basagni2006new}, which greedily selects the site with the largest residual energy of surrounding sensors to visit. To get a decent solution within tolerable consuming time, Liang~\emph{et al.} \cite{liang2010prolonging} present a three-stage heuristic method that iteratively does local improvements on a distance-constrained shortest path problem that approximates the original lifetime maximization problem. Meta-heuristic methods are also utilized to solve the lifetime maximization problem, they are capable of finding near-optimal solutions by exploring the space of possible solutions efficiently. Nonetheless, they can be very time-consuming when facing large-scale problems, owing to the repeated evaluations of the objective function.
Some existing methods introduce ant colony optimization~(ACO) algorithm~\cite{zhong2012ant,wang2018improved,maheshwari2021energy} to improve the lifetime of WSNs. 
In recent years, hyper-heuristic algorithms also have been leveraged to solve the combinational optimization problem in WSNs. A hyper-heuristic framework is proposed in~\cite{zhong2019hyper}, which can automatically construct a solution to optimize the moving path of the sink to prolong the lifespan of WSNs based on several heuristic rules. While the above methods are highly dependent on the predefined heuristics. There is no guarantee that human knowledge is necessarily optimal.

Avoiding the design of complicated computing mechanisms and massive feature engineering, RL-based methods are applied in WSNs to increase the lifetime. 
Many works focus on improving energy efficiency and prolonging the lifetime of cluster-based WSNs, in which RL methods are mainly applied to select cluster heads and decide the visit of the sink to collect data. For the clustering or the election of cluster head, the tabular Q-learning method is leveraged in~\cite{forster2009clique,mustapha2017energy,soni2018novel} to learn the optimal cluster head to address energy challenges in WSNs. As for finding the shortest path to collect data, the tabular Q-learning method is also used in~\cite{krishnan2021reinforcement} to improve the movement route of the mobile sink reaching the cluster heads to collect data. 
However, the state space covered by Q-table is limited in the above works, which results in a lack of scalability and can not be used in dynamic networks. In this paper, we propose a deep reinforcement learning approach, which is able to handle more complex WSNs with vast state space and dynamic situations. Besides, in contrast to the above methods for cluster-based protocol WSNs, we focus on the problem of scheduling the traversal path of the sink to maximize the lifetime in an instance model of flat-based protocol WSNs, which is more suitable for the network with low redundancy signals.

\section{Problem Definition}\label{section3}

This section describes the process of transmitting data from the sensor to the sink in WSNs, using the notations provided in Table~\ref{notaions}. We then present the mathematical formulation of the problem of maximizing the lifetime of WSNs.

\begin{table}[htbp]
	\centering
	\caption{Problem notation}\label{notaions}

		\begin{tabular}{>{\centering\arraybackslash}m{0.3\columnwidth}|>{\centering\arraybackslash}m{0.6\columnwidth}}
			\toprule
			\textbf{Symbol} & \textbf{Definition}  \\
			
			\hline
			$L=\{l_1,\cdots,l_n\}$ & The set of sites \\
			\hline
			$N=\{n_1,\cdots,n_m\}$ & The set of sensor nodes \\
			\hline
			$\alpha$ & The sink \\
			\hline
			$T$ & The lifetime of the WSN \\
			\hline
			$z_i$ & The sensing data rate of $n_i$ \\
			\hline
			$d_{max}$ & The maximum transmission range of sensor nodes\\
			\hline
			$d_{i,j}$ & The euclidean distance between node $i$ and node $j$\\
			\hline
			$t_j$ & The sojourn time of the sink stays at $l_j$ \\
			\hline
			$\Gamma_j^k$ & Whether the sink settles in site $l_j$ in $k$th round \\
			\hline
			$E_i$ & The initial energy of $n_i$ \\
			\hline
			$u_i^k$ & The residual energy of sensor $n_i$ in $k$th round \\
			\hline
			$et_i^k$ & The consumption rate of sending one bit of data for $n_i$ in $k$th round \\
			\hline
			$er_i^k$ & The consumption rate of receiving one bit of data for $n_i$ in $k$th round \\
			\hline
			$f_i^k$ & The amount of data sent by $n_i$ in $k$th round \\
			\hline
			$g_i^k$ & The amount of data received by $n_i$ in $k$th round \\
			
			\bottomrule
		\end{tabular}
	
\end{table}

\begin{figure}[htbp]
	\centering
	\includegraphics[scale=0.2]{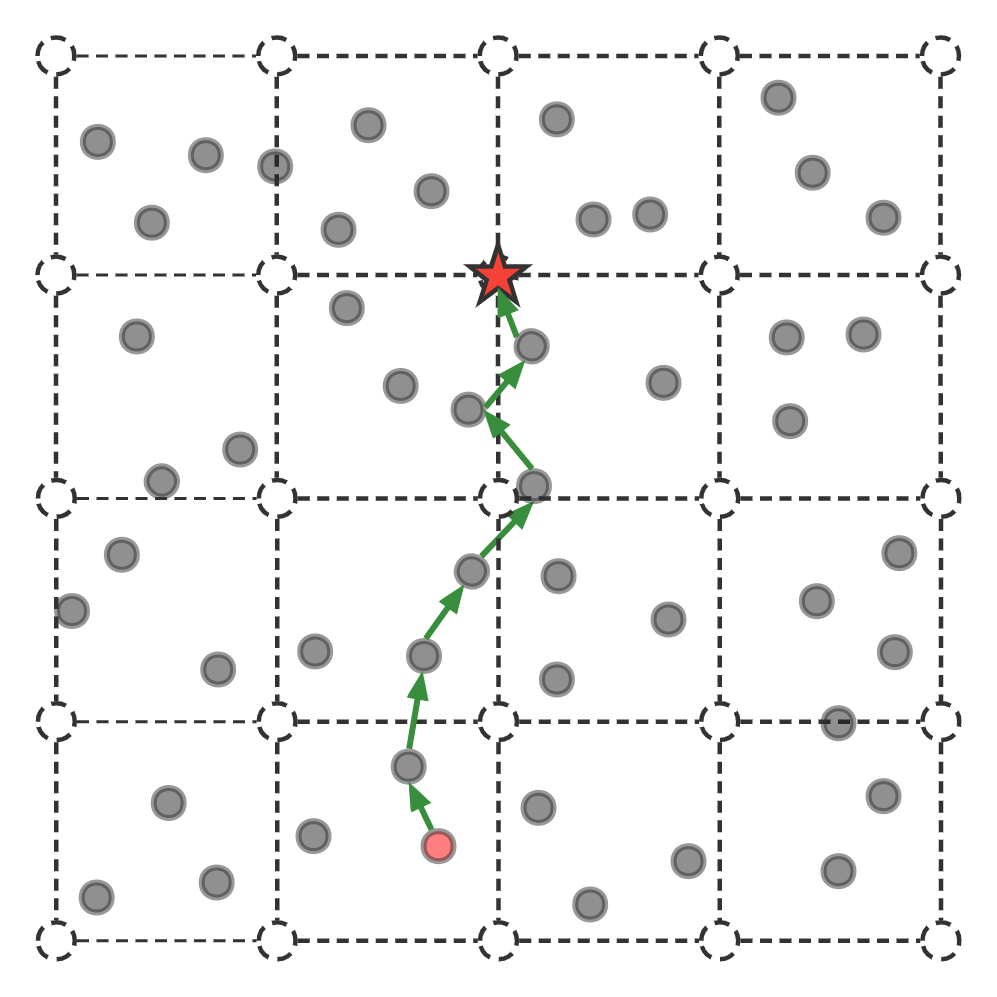}
	\caption{The forward route of data transmission from the source sensor to the sink. The dashed circle denotes the sites, the red five-pointed star refers to the sink, and the solid circle represents the sensor nodes. The solid red circle refers to the source sensor, and the arrow represents the data flow of two nodes within the transmission range. }\label{sensor_sink}
	\vspace{-3mm}
\end{figure}

A WSN consists of a sink $\alpha$ which can stop in a set of locations (i.e., sites) $L=\{l_1,\cdots,l_n\}$, and a large set of homogeneous sensor nodes $N=\{n_1,\cdots,n_m\}$ scattered in a geographic area following the sensor-updating function $f$. The distribution of sensors in $k+1$ round can be defined as $N^{k+1}=f(N^k)$. In particular, the distribution of sensors in the static sensor network does not change over time, i.e. $N^{k}=N^{k+1}=f(N^k)$. Both the sensor nodes and the sink have a maximum transmission range $d_{max}$. Each sensor node $n_i$ is assumed to generate information at a fixed rate of $z_i$ during its life span. Hence, the data produced by $n_i$ in the time interval $\Delta t$~(the minimum time of each round) can be denoted as $z_i * \Delta t$. Then the data will be forwarded to the sink node in one hop if the Euclidean distance $d_{i,\alpha}$  between the sensor node $n_i$ with the sink $\alpha$ is smaller than $d_{max}$, otherwise, through a multi-hop transmission tree if $d_{i,\alpha} > d_{max}$. The data transmission from the sensor node to the sink in WSN is shown in Fig.~\ref{sensor_sink}. 

At each round $k$, the sink reaches one of the possible sites in $L$ or does not move~(i.e., stays at the location in the previous round $k-1$). The traveling time $\dot{t}$ between two sites is ignored, as $\dot{t} \ll \Delta t$. When the sink reaches a new site or leaves the previous site, a data packet containing the current location of the sink will be sent to all sensors to make them aware of the new site of the sink. We denote the sojourning time of the sink at each location $l_j$ as $t_j$. Therefore, the lifetime $T$ of the WSN can be written as follows:
\begin{equation}
	T=\sum_{j=1}^{n} t_j \nonumber
\end{equation}
Where the $t_j=\sum_k  \Gamma_j^k$. $\Gamma_j^k$ is a binary variable that denotes whether the sink settles in site $l_j$ at the $k$th round: if the sink stays at site $l_j$, the $\Gamma_j^k$ will be 1, otherwise, it is 0. For the case of WSN with a single sink, only one site places the sink at each round $k$: $\sum_{j=1}^m \Gamma_j^k=1$.  

In the beginning, each sensor node $n_i$ has the initial energy $E_i$ and the WSN dies when it occurs one of the sensors exhausts its energy. The data flow of a sensor node involves sending data generated by itself and the transmission of data comes from other nodes $j$ ($d_{i,j}<d_{max}$). The incoming data flow at $n_i$ consists of data sent by other nodes, and the outgoing data flow includes sending its data and the data originating from other sensors.

Both the process of sending or receiving data of sensor node  $n_i$  require energy. The transmission power consumption is closely coupled with the route selection. The consumption rates of sending and receiving data of sensor $n_i$ in $k$th round can be denoted as ${et_i^k}$  and $er_i^k$. We consider the $er_i^k$, $\forall k$, is a constant and the consumption rate $et_{i,j}^k$ for data sending from sensor $i$ to sensor~(or the sink) $j$ is proportional to the Euclidean distance $d_{i,j}$, as it's shown in \eqref{consumtion rate}. The $a$ and $b$ are constant coefficients related to the transmission media properties.
\begin{equation}\label{consumtion rate}
	et_{i,j}^k=a \times d_{i,j}^{2} + b
\end{equation}	

The multi-hop transmission tree $H$ is leveraged to decide the data transmission routes between sensors and the sink, which can be constructed by the Flow Augmentation algorithm~(FA)~\cite{chang2004maximum}. $H$ represents the shortest cost path routing from the origin node~(i.e. sensor) to the destination node~(i.e. sink) through computing link costs reflecting both the communication energy consumption rates and the residual energy of the two end nodes. That is to say, the FA algorithm takes the distribution of sensors, the residual energy of sensors, and the location of the sink as input, and outputs the amount of data sent $f_i^k$~(or received $g_i^k$) by sensor $n_i$ in the $k$th round, which can be formulated as:
\begin{equation}\label{consumtion data}
	[f_i^k,g_i^k]=\text{FA} \ (N^k,loc(\alpha)^k, U^k)
\end{equation}	
Where the $U^k=[u_1^k,u_2^k,\cdots,u_m^k]$ represents the residual energy of sensors, and the $loc(\alpha)^k$ denotes the site of the sink stays at $k$th round.

Based on \eqref{consumtion rate} and \eqref{consumtion data}, the energy consumption rate per unit time of sensor $n_i$ at $k$th round is $et_i^k \cdot f_i^k + er_i^k \cdot g_i^k$. Thus, the energy cost of each sensor $n_i$ during its lifetime in the WSN can be defined as follows:
\begin{equation}
	\sum_k( et_i^k \cdot f_i^k + er_i^k \cdot g_i^k) \nonumber
\end{equation}

Following the definition
in~\cite{zhong2019hyper}, the problem of maximizing the lifetime of the WSN can be formulated as a MILP model:
\begin{align}
	max~ &T=\sum_{j=1}^{n} t_j \\
	s.t.~ &t_j=\sum_k  \Gamma_j^k, l_j\in L \\
	&\sum_{j=1}^m \Gamma_j^k=1 \\
	&\sum_k( et_i^k \cdot f_i^k + er_i^k \cdot g_i^k)-E_i \geqslant 0,  \forall n_i\in N \label{energy constraint}\\
	&g_i^k+z_i * \Delta t=f_i^k , \forall n_i\in N \label{data flow constraint}
\end{align}
Equation \eqref{energy constraint} is an energy constraint that represents that the total energy cost of each node during its lifetime is less than the initial energy. Besides, the data flow conservation \eqref{data flow constraint} ensures that the amount of incoming data with the data produced by the sensor $n_i$ equals the amount of its outflow data.

\section{Proposed Method}\label{section4}
In this section, we illustrate our framework formally, including the design of MDP, the structure of the neural network for evaluating the agent's state-action value, and the training method. Firstly, the overall framework of our method is introduced. Secondly, we formulate the movement process of the sink to maximize WSN's lifetime as a MDP. Then, we represent the WSN as a weighted undirected heterogeneous graph. In addition, the way to learn the representation of nodes in WSN based on the GNN and multi-head attention is given. Finally, we introduce the training algorithm in detail.
\subsection{Overview}

\begin{figure*}[t]
	\centering
	\includegraphics[scale=0.32]{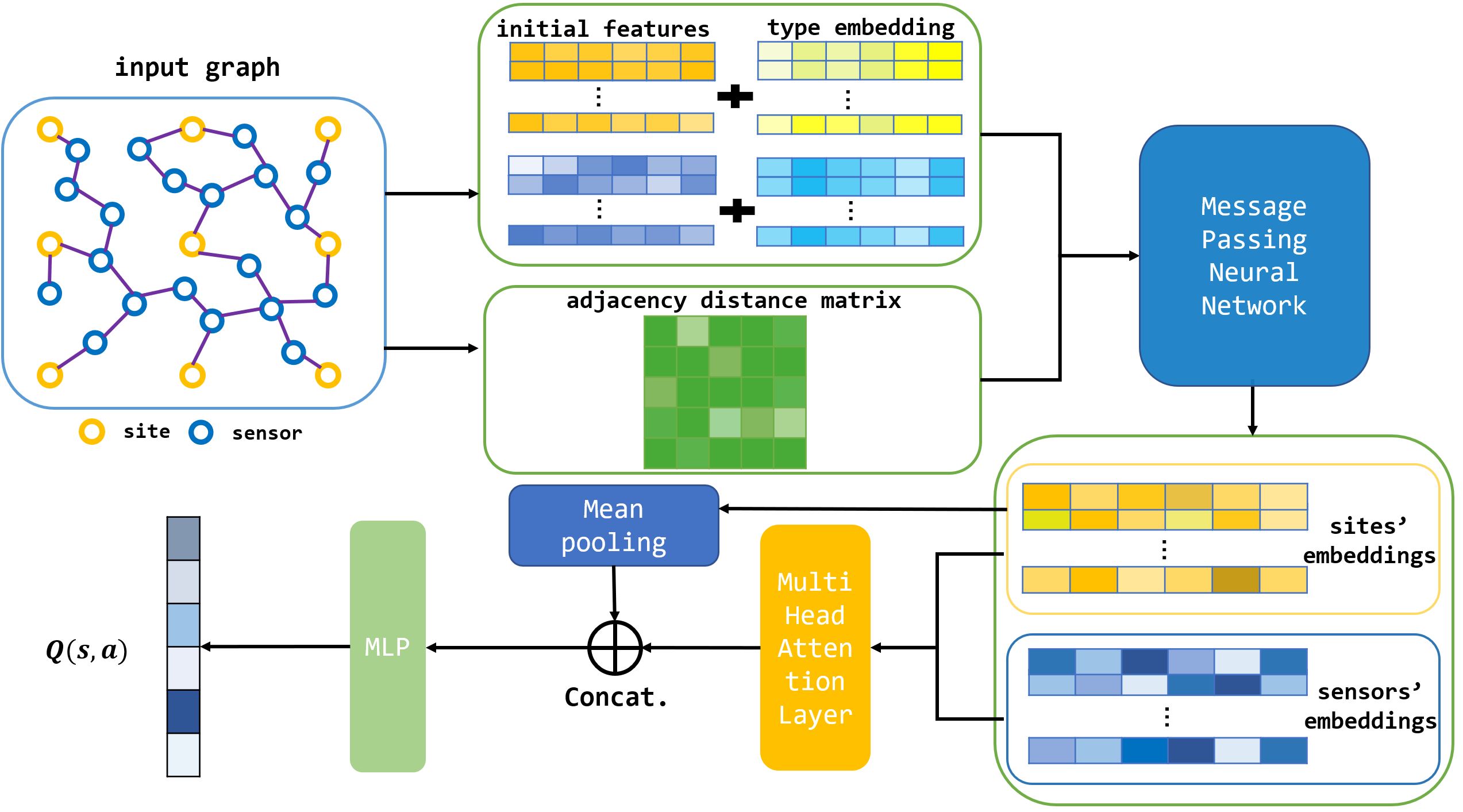}
	\caption{The overall framework of HGFF }\label{framework}
\end{figure*}

The lifetime maximization with mobile sink problem of different WSNs have the same structure but differ in the geographic area, distribution, and sensor-updating function. Due to the requirement for high-quality human knowledge, it's challenging to design optimal heuristic rules to solve the problem. Different from previous work mainly using heuristics, a deep reinforcement learning-based framework HGFF is proposed to learn the optimal movement strategy for the sink in this paper. 

Firstly, we formalize the WSN as a weighted undirected heterogeneous graph. Due to the different roles and attribute information, the sensors and sites are naturally seen as two types of nodes in the heterogeneous graph. Then, the learnable node type embedding is incorporated with the graph neural network to learn the representations of sensors and possible sites the sink visits with the relationship between them. To further enhance representations, we introduce the multi-head attention to "fuse" features, which means learning and leveraging the compatibility between embeddings of the two kinds of nodes. Different from the attention in common graph attention networks, we focus on all the sensor nodes in the WSN graph instead of only the neighbor nodes. Such a method is more suitable for dynamic WSNs, in which the surrounding sensors are changing at each time step.

Based on the Double Deep Q-learning (DQN) algorithm~\cite{van2016deep}, the agent learns the policy of guiding movement of the sink to maximize the network's lifetime from thousands of training experiences. The overall framework of the agent's Q-network is shown in Fig.~\ref{framework}. The Q-network takes the state of the virtual WSN environment, which is represented as graph data, and outputs the evaluation of the value for agent stays at each site: $Q(s,a)$. To learn the heterogeneity of nodes, the representation of nodes will be enhanced by combining learnable type embeddings explicitly in aggregating information from neighbors and combining the information to update representations in the next iteration. Building on the updated embeddings in graph learning, we aim to learn the compatibilities of sites with sensor nodes to improve the representations, which are the bases for estimating the Q-value. The estimated Q-value is updated during trial-and-error and will be greedily utilized to obtain the optimal moving strategy in the inference of the model.

\subsection{Markov Decision Process Formulation}
We formalize the moving process of sink in a WSN as a Markov Decision Process~(MDP). The purpose of such modeling is to address the problem: determining the route for the mobile sink, together with the sojourn times at which the sink stays on each site, so as to maximize the lifetime of the WSN. The MDP can be defined as a five-tuple $\langle \mathcal{ S, A, T, R, \gamma} \rangle$. $S$ represents the environmental information and the possible configurations of the WSN environment. $A$ denotes the available action set of the agent and $T$ defines the probability distribution over possible next states. $\gamma \in [0,1]$ is  the discounted factor. It is worth mentioning that the Markov property is satisfied in the virtual WSN environment. Given the present state of the WSN environment, the future decision of the sink~(i.e. select a site to move) is independent of the past. In other words, the state of the WSN environment includes all the relevant information about its history: $\mathbb{P}[S_{t+1}|S_t]=\mathbb{P}[S_{t+1}|S_1,\cdots, S_t]$.

Based on the MDP, we simulate the decision process of which site the sink stays at each time step~(i.e. $\Gamma_j^k$) until the overall WSN dies. The main parts of MDP are defined as follows:

\textbf{States:} The state for the agent includes all the attribute information of the sensor nodes and sites, which is composed of the following global information:
\begin{enumerate}
	\item Property of node: sensor, site, and if it belongs to the set of sites, whether there is a sink on this site.
	\item Current coordinate position~(including x-axis and y-axis) of the sink, sensors, and sites.
	\item Residual energy of the sensor nodes.
	\item Energy consumption per unit time of sensor nodes.
	\item Euclidean distance $d_{i,j}$ between every two nodes if $d_{i,j}<d_{max}, i,j\in L \cup N$ .
\end{enumerate}

\textbf{Actions:} Since the time for the sink to move between the site nodes is ignored, we assume that any one of the sites is available to the sink at each time step. Thus, the action for the agent is to select a site from the set of sites, and the length of action space is equivalent to the range of site set $L$: $\mid{A}\mid=\mid{L}\mid$

\textbf{Rewards:} In fact, maximizing the WSN's lifetime $T$ is equivalent to maximizing the accumulated sojourn times of the sink at the sites, and also equivalent to maximizing the expected episode's length of the agent. Intuitively, we define the reward as a constant: $r_t=1$, for each time step $t$. The goal of the agent is to learn the optimal movement policy $\pi$ for the sink to maximize the expected return with a discount factor $\gamma$: $E_{\pi} \{\sum_{t=0}^\infty \gamma^t r_t\}$. With $r_t=1$, the goal of the agent can also be written as  $max~ E_{\pi} \{\sum_{t=0}^T\ \gamma^t \}$.

\textbf{Transitions:} At each time step, after the agent makes the decision to choose one site to visit, the environment will process the decision to step into the next state definitely: $ \mathcal{ T: S \times A \rightarrow S}$. The energy consumption of each sensor node due to sending and receiving data will be calculated to obtain the new state.

\textbf{Terminal Criterion:} When the residual energy $u_i \leqslant 0, \forall i \in N$ happens, the episode ends.

\subsection{Learning over graphs}

\subsubsection{Graph Representation Formulation}\label{graph_formulation}

We define the WSN as a weighted heterogeneous graph: $G=(V,E,w,\phi,\psi)$, where the set of nodes $V$ is constituted of two kinds of nodes: sensors in set $N$ and sites in set $L$. The set of edges $E$ represents the connection between two nodes that are able to communicate with each other~(i.e. they are within the transmission range $d_{max}$). $w$ refers to a function mapping undirected edges to their weight values: $E\rightarrow W$, where $W=\left\{W_{e}:\forall e \in E \right\}$ means the set of weights. $\phi$ is the node type mapping function: $V \rightarrow T_v$, where $T_v=\left\{\phi(v):\forall v \in V\right\}$ denotes the set of possible node types. $\psi$ is the edge type mapping function: $E \rightarrow T_e$, where $T_e=\left\{\psi(e):\forall e \in E\right\}$ denotes the set of possible edge types. In the graph $G$, there are two types of nodes and one type of edges: $|T_v|=2$, $|T_e|=1 $. It is worth noting that $G$ must be a fully connected graph, otherwise, there will be isolated nodes that cannot send data to the outside. 

We design 5 features for sensor node $n_i$ at each time step $k$: the mark for the type of node, the x-coordinate of the node, the y-coordinate of the node, the ratio of the residual energy to the initial energy in  $k$th time step: $u_i^k/E_i$, and the energy consumption per unit time in $k$th time step: $et_i^k \cdot p_i^k + er_i^k \cdot q_i^k$. For the features of site $l_j$, they are the mark for the type of node, the x-coordinate of the node, and the y-coordinate of the node. Note that the elements for both features are normalized, except for the identification mark for different nodes.

To include the distance information between different nodes in the representation, we explicitly define the weight $W_{e_{ij}}$ of connection between node $i$ and node $j$ be the normalized Euclidean distance of two nodes:
\begin{align}
	w(e_{ij})=normlize(d_{ij}) \nonumber
\end{align}
The information on the above nodes and edges is used as the initial feature to train the graph neural network in the next section. 

\subsubsection{Message Passing Neural Networks with learnable Node-type embedding}\label{GNN}
To leverage type information to improve training, we incorporate node type embedding into the original Message Passing Neural Network~(MPNN)~\cite{gilmer2017neural}, which is a general learning framework used widely in many famous graph networks with specific implementations. Based on the input node feature $\mathbf{x}_v\in\mathbb{R}^d_0$~(d=5) and the structure of graph $G$ constructed in sec~\ref{graph_formulation}, we aim to learn an $d_h$~(we use $d_h=64$) dimensional representation vector $\mathbf{h}_v^l$ for each vertex $v\in V$, where $l$ denotes the current iteration or network layer. Firstly, the input feature $\mathbf{x}_v$ is input to a linear projection with parameters $W^x$ and $B^x$: $\bm{\mu}_v^0=W^x\bm{x}_v+B^x$. Then  $\mathcal{L}$ rounds of message passing will be leveraged to update the embeddings, according to \eqref{message} and \eqref{update}.

Although the original MPNN can be used to model homogeneous graphs, the omission of node or edge types could make it suboptimal for heterogeneous graphs. To emphasize the heterogeneity between two kinds of nodes, we allocate an embedding $\bm{t}_{\phi(v)}^l$ for each type of vertex $\phi(v), \forall v\in V$, where the contents of $\bm{t}_{\phi(v)}^l$ are adjusted during the training process. Therefore, the node embedding $\bm{h}_v^l$ consists of two parts: node embedding $\bm{\mu}_v^l$ in original MPNN and learnable node type embedding $\bm{t}_{\phi(v)}^l$, where the $\phi(v)$ is the type of node $v$. The embedding $\bm{h}_v^l$ of node $v$ in iteration $l$ is defined as $\bm{h}_v^l=\bm{\mu}_v^l||\bm{t}_{\phi(v)}^l$, where the $||$ denotes concatenation operation.

Based on the aggregation from the neighbor nodes, the graph neural network updates node representation iteratively following:
\begin{align}
	&\bm{m}_v^{l+1}=M_l(\bm{h}_v^l,\left\{\bm{h}_u^l\right\}_{u\in N(v)},\left\{\omega_{uv}\right\}_{u\in N(v)}) \label{message} \\
	&\bm{h}_v^{l+1}=U_l(\bm{h}_v^l,\bm{m}_v^{l+1})\label{update}
\end{align}
Where $M_l$ denotes the message function and $U_l$ represents the update function. The message funtion aggregates information from neighborhood nodes $\left\{\bm{h}_u^l\right\}_{u\in N(v)}$ with adjacent edges $\left\{\omega_{uv}\right\}_{u\in N(v)}$, and pass on the information to the next layers. The node representation is updated iteratively through the non-linear transformation in the update function.

\subsection{Attention-based feature fusion}

Aggregating the embeddings from neighbors, the GNN in Section~\ref{GNN} learns the representations of sites and sensor nodes: $\{\bm{h}_v^\mathcal{L}\}=\{\bm{h}_l^\mathcal{L}\} \cup \{\bm{h}_n^\mathcal{L}\}$, where $v\in V, l\in L, n\in N$. To further learn the association between nodes and enhance the model's expressive capacity, we adopt the attention mechanism~\cite{vaswani2017attention}, which requires a query and a collection of key-value pairs to map the output. Formally, we define the queries consist of site nodes' embeddings $\{\bm{h}_l^\mathcal{L}\}$ and keys and values $\{\bm{h}_n^\mathcal{L}\}$ come from the sensor nodes' embeddings. The queries, keys, and values are calculated by the linear transformations of the sites' embeddings and sensors' embeddings respectively:
\begin{align}
	\bm{q_i}=W_q*\bm{h_l}^\mathcal{L}, \bm{k_j}=W_k*\bm{h_n}^\mathcal{L}, \bm{v_j}=W_v*\bm{h_n}^\mathcal{L} \nonumber
\end{align}
where $\forall i\in L, \forall j \in N$, 
parameter $W_q$ is learnable matrice~($d_q*d_h$), and $W_k,W_v $ have the same dimension $d_k*d_h$. The compatibility of the query $\bm{q}_i$ of site node $i$ with the key $\bm{k}_j$ of sensor node$j$: $u_{ij}\in \mathbb{R}$ is computed following the dot-product in \eqref{dot production}.

\begin{align}\label{dot production}
	u_{ij}=\frac{\bm{q}_i^T\bm{k}_j}{\sqrt{d_k}}
\end{align}

Based on the compatibilities $u_{ij}$, the attention weights $a_{ij}\in [0,1]$ can be computed by the softmax function:
\begin{align}
	a_{ij}=\frac{e^{u_{ij}}}{\sum_j e^{u_{ij}}} \nonumber
\end{align}
Then a convex combination of \bm{$v_j$} is used to produce vector $h_i^{'}$:
\begin{align}\label{value}
	\bm{h}_i^{'}=\sum a_{ij}\bm{v}_j \nonumber
\end{align}

Due to the stability that comes from learning different types of messages from neighbors, multi-head attention with $M=8$ heads is leveraged to compute the value in \eqref{value} repeatedly with M different learnable matrices. After calculating various attention maps, we combine all the learned representations. Specifically, the dimensionality of query or key is $\frac{d_h}{M}$. The resulting vectors are denoted as $\bm{h}_{im}^{'}$ for $m\in \{1,...,M\} $ and learnable matrices $W_m^O$ are used to project back the vectors to $d_h$ dimension vector. Finally, the multi-head attention value $\bm{h}_i^*$ for site node $i$ can be defined as following:
\begin{align}
	\bm{h}_i^*=\sum_{m=1}^M W_m^O\bm{h}_{im}^{'}
\end{align}

Selectively focusing on the most important parts of the sensors' embeddings following the above way, the site node embeddings $\{\bm{h}_i^*\}_{i\in L}$ integrated the global sensor information are obtained. Further, the mean pooling of sites' embeddings $\frac{\sum_{i=1}^L\bm{h}_i^*}{L}$ is concatenated with $\{\bm{h}_i^*\}_{i\in L}$, and then a Multi-Layer Perceptron~(MLP) function is leveraged to output the final evaluate the value of site nodes that the agent may visit at time step $t$: $Q(s_t,a_t)$.

\subsection{Training method}\label{DDQN}

The Double DQN algorithm~\cite{van2016deep} is employed to learn the optimal policy for planning the traversal path of the sink. Untangling the selection and evaluation in the update of the Q value, the Double DQN method reduces the overestimation error by learning two value functions, one of which is used to decide the greedy policy and the other to decide the value. The learning target $Y_t$ of Double DQN algorithm at $t$ time step is shown in \eqref{target of ddqn}:
\begin{align}\label{target of ddqn}
	Y_t=R_{t+1}+\gamma Q(s_{t+1},argmax_aQ(s_{t+1},a;\theta_t),\theta_t^-)
\end{align}
Where $\theta_t$ refers to the parameters of the Q-network, $\theta_t^-$ are the parameters of the target network, and $\gamma$ refers to the discount factor. The parameters of Q-network $\theta$ is updated following:
\begin{align}\label{update network}
	\theta_{t+1}\leftarrow \theta_t +\alpha(Y_t-Q(s_t,a_t;\theta_t)) \nabla_{\theta_t}Q(s_t,a_t; \theta_t) 
\end{align}
where $\alpha$ is a scalar step size. The pseudocode of the training method is described in Algorithm~\ref{DDQN_algorithm}.

\begin{algorithm}[htbp]
	\caption{Double DQN training algorithm}\label{DDQN_algorithm}
	\begin{algorithmic}[1]
		\STATE Initialize the replay buffer  $\mathcal{D}$ and network $Q_{\theta}$ with random weights $\theta$ and target network $Q_{\theta^-}$ with $\theta^-=\theta$
		\FOR{episode 1 to $\mathcal{M}$}
		\WHILE{not done}
		\STATE Get current state $s_t$ 
		\STATE Compute $Q$ values according to network model $Q_{\theta}$ based on $s_t$ and select an action $a$ following \emph{epsilon greedy}:
		\STATE $$a=
		\begin{cases}
			argmax_aQ_\theta(s_{t+1},a) & \text{with prob $1-\epsilon$}\\
			\text{sample from action space}  & \text{with prob $\epsilon$}
		\end{cases}$$
		
		\STATE Execute $a$ in the environment
		
		\STATE Get reward $r_t$ and next state $s_{t+1}$
		
		\STATE Store transition ($s_t,a,r_t,s_{t+1}$) in $\mathcal{D}$
		
		\STATE Sample mini-batch experience from $\mathcal{D}$
		\STATE For the sampled experience, compute the update target in \eqref{target of ddqn} and update the parameters $\theta$ according to \eqref{update network}
		
		\ENDWHILE
		\STATE Every $C$ steps perform $\theta^-=\theta$
		\ENDFOR
		
	\end{algorithmic}
\end{algorithm}

\section{Experiments}\label{section5}

In this section, we introduce the experimental setup and results of our proposed method. In the experimental setup, the designed WSN environments for evaluating different approaches are introduced. Then, we present the indicators to probe the performance of different algorithms. Moreover, the baseline methods and their hyper-parameters are illustrated in detail. Then, the hyper-parameters during the training of our method and the baseline DRL-based method are given. In the experimental results, we show the results of the conducted extensive experiments on both static and dynamic maps, as well as some analysis. In addition, we further evaluate the effectiveness of the component of our method via ablation studies. What's more, the visualization of the sink mobile path is given, which directly reflects the effectiveness of our method.

\subsection{Experimental Setup}

\subsubsection{Design of Virtual Environments}\label{sec_5_1_1}
Two kinds of virtual environments are designed to simulate WSNs: stationary and dynamic environments. In stationary environments, the deployment of sensor nodes does not change during the lifetime of WSNs, and the energy load balancing of the network only depends on the movement of the mobile sink. Further, to cater to most situations in the real-world, dynamic WSN environments are also designed to test our proposed method, that is, during the operation of the network, the deployment of sensor nodes will change over time. The dynamic environments pose a great challenge to the response time of the algorithm. Besides, the initial location distribution of sensors is randomly generated.

\begin{table}[htbp]
	\centering
	\caption{Ten types of maps, which differ in the geographic area, the number of sensors and sites}\label{WSN environments}
		\begin{tabular}{{>{\centering\arraybackslash}m{0.2\columnwidth}|>{\centering\arraybackslash}m{0.2\columnwidth}|>{\centering\arraybackslash}m{0.2\columnwidth}|>{\centering\arraybackslash}m{0.2\columnwidth}}}
			
			\toprule
			\textbf{Index} & \textbf{Sensors} & \textbf{Sites} & \textbf{Mapsize} \\
			
			\hline
			1 & 30 & 5$\times$5 & 100$\times$100 \\
			\hline
			2 & 50 & 5$\times$5 & 100$\times$100 \\
			\hline
			3 & 100 & 5$\times$5 & 100$\times$100 \\
			\hline
			4 & 100 & 10$\times$10 & 150$\times$150 \\
			\hline
			5 & 200 & 5$\times$5 & 100$\times$100 \\
			\hline
			6 & 200 & 10$\times$10 & 150$\times$150\\
			\hline
			7 & 100 & 5$\times$15 & 50$\times$150 \\
			\hline
			8 & 100 & 10$\times$10 & 100$\times$100\\
			\hline
			9 & 300 & 10$\times$10 & 150$\times$150\\
			\hline
			10 & 500 & 20$\times$20 & 150$\times$150 \\ 
			\bottomrule
		\end{tabular}
\end{table}

Specifically, ten types of WSN environments are designed for both static and dynamic cases respectively. The initial location distribution of sensors in each type of dynamic map is consistent with that in the corresponding static map.
All the scenarios are listed in Table~\ref{WSN environments},  which differ in map size and the number of sensors and sites. In particular, map 7 refers to the sensors and sites distributed in a rectangle sensing region (50*150). Besides, we set half of the sites randomly to be inaccessible in map 8 to simulate accidental situations in the real world.  For other maps, the complexity is growing with the number of sensors and sites gradually increasing and distributed on a larger map. The more complex maps bring greater challenges to the performance of the algorithm and computing resources.

To validate the performance of methods, we generate ten maps for each type of WSN environment respectively. As described in Section~\ref{section3}, the deployment of sensors is randomly generated and the locations of sites are arranged in squares. In each map, the initial 2D coordinate of every sensor is sampled uniformly at random in the square of size ~$[0,mapsize] \times [0,mapsize]$. In dynamic environments, the coordinates of sensors are changed at each time step. To simulate such a dynamic of sensors, the coordinate of the sensor $i$ at time step $t$ is determined by sampling a normal distribution $\mathcal{Z}$, in which the mean is the initial location of sensor $i$ and the variance is set to 3. For the routing of data from sensors to the sink, the FA algorithm~\cite{chang2004maximum} is leveraged to construct the multi-hop flow routing tree, and the parameters are as follows. In both the stationary map and dynamic map, the maximum transmission range $d_{max}$ is set as 30 meters, and the constant coefficients $a,b$ for computing the energy consumption of sending data in \eqref{consumtion rate} are set to 50 $nJ/bit$ and 100 $pJ/bit/m^{2}$ respectively. The sensing rate $z_i$ is 1bit/s and the least sojourn time $\Delta t$ is 3600 seconds.

\subsubsection{Evaluation Metrics}

The {\itshape average lifetime} and {\itshape computation time} are leveraged as indicators to evaluate the quality of methods. The {\itshape average lifetime} intuitively reflects the optimality of the solution, which can be denoted as $\overline{T}$. The {\itshape computation time} $C_t$ refers to the time elapsed for the sink to make the decision of staying at which site next timestep, in seconds, which reflects the efficiency of methods to obtain solutions. The calculation of $\overline{T}$ and $C_t$ is defined as follows:
\begin{align}
	\overline{T}=\frac{\sum_{i=0}^n T_i}{n} \nonumber \\
	C_t=\frac{time_{tot}}{T} \nonumber
\end{align}
Where $T_i$ represents the lifetime of the WSN in $i$ episode, $time_{tot}$ means the running time of an episode from resetting the environment until the WSN dies, and the $T$ defined in Section~\ref{section3} means the cumulative decision count of sink in an episode. Note that the corresponding actual lifetime of the WSN in $i$ episode is $T_i \cdot \Delta t$ seconds. For the proposed method and RL-based baseline methods, the two test indicators are measured by averaging the results of 20 test episodes after training, in which the agent performs action selection greedily. The inference time of the proposed method and all the compared methods are evaluated on a workstation with Intel(R) Core(TM) i9-10900K CPU @ 3.70GHz.

\subsubsection{Methods for Comparison}
In our experiments, several existing algorithms are selected for comparison, which can represent three categories of methods: heuristic methods, hyper-heuristic methods, and deep reinforcement learning methods.

\textbf{Heuristic methods:} For the heuristic methods, a rule-based method: Greedy Maximum Residual Energy (GMRE) method~\cite{basagni2006new} is reproduced in our experiments, which leads the sink to move to the site node with the most residual energy greedily. In addition, the ACO method~\cite{dorigo2018introduction} is also implemented in the experiments. With relatively strong global search ability, the ACO algorithm has been shown able to get near-optimal solutions for the lifetime maximization problem in WSNs~\cite{zhong2012ant,maheshwari2021energy}. The details of the two methods are as follows:
\begin{enumerate}
	\item \textbf{GMRE:} Intuitively, GMRE proposed a heuristic $\zeta$ that the sink moves to the site with the largest residual energy of the sensors within a certain distance  $\mathcal{L}$:
	\begin{align}
		\zeta=max~u_i^t,\forall l_j\in L,n_i\in N, d(l_j,n_i)<\mathcal{L} \nonumber
	\end{align}
	Following the rule $\zeta$, the site which covers sensors with maximal residual energy is selected.
	
	\item \textbf{ACO:}  In nature, real ants can find the shortest path from a food source to a nest by leveraging pheromones without vision. Inspired by real ants, the ACO algorithm~\cite{dorigo2018introduction} is proposed to solve combinational optimization problems. In the ACO algorithm, there is a group of artificial ants to search for the optimal solutions according to the heuristic information and global pheromones. The heuristic information is predefined manually before search and the pheromones are updated by the results of the historical exploration of ants. In this paper, the heuristic functions in \cite{zhong2012ant} are used to compute the initial pheromone and update the selection probability of each ant, including \textit{average communication hop}, \textit{average residual energy} and \textit{maximum step distance}. In this paper, the experimental settings of the ACO algorithm are listed in Table~\ref{ACO parameters}.

	\begin{table}[htbp]
		\centering
		\caption{Hyper-parameters for the ACO algorithm}\label{ACO parameters}
		\resizebox*{0.9\linewidth}{!}{
			\begin{tabular}{>{\centering\arraybackslash} c | c }
				\toprule
				
				\textbf{Parameter} & \textbf{Value}\\
				
				\hline
				Number of artificial ants & 10  \\
				\hline
				Exploitation rate to construct solution & 0.95  \\
				\hline
				Pheromone reinforcement rate & 0.5  \\
				\hline
				Weight of heuristic information & 0.1  \\
				
				\hline
				Weight of the average communication hops & 10 \\

				\bottomrule
			\end{tabular}
		}
	\end{table}

\end{enumerate}

\textbf{Hyper-heuristic methods:} Searching for the optimal solution among different heuristics automatically, the hyper-heuristic methods are proposed to improve performance and flexibility. In this paper, a self-learning gene expression
programming~(SL-GEP) algorithm~\cite{zhong2019hyper} is selected as one of the baselines to compare with our proposed method. In SL-GEP, the chromosome representation of an individual consists of subfunctions, and the high-level heuristic is obtained by the combination of these functions. Besides, the subfunctions are made up of linking functions and the Automatically Defined Functions(ADFs), which are the combination of handcraft low-level heuristic functions. The population is evolved by the use of evolutionary operations, such as mutation, crossover, and selection based on a modified differential evolution method to obtain individuals with higher fitness values. In our experiments, the experimental settings for  SL-GEP are shown in Table~\ref{SLGEP parameters}.

\begin{table}[htbp]
	\centering
	\caption{Hyper-parameters for the SL-GEP algorithm}\label{SLGEP parameters}
		\resizebox*{0.9\linewidth}{!}{
		\begin{tabular}{>{\centering\arraybackslash} c | c }
			\toprule
			
			\textbf{Parameter} & \textbf{Value}\\
			
			\hline
			Evolving generations & 500  \\
			\hline
			Population size & 10  \\
			\hline
			Length of the head in main program & 10  \\
			\hline
			Length of head in ADFs & 3  \\
			\hline
			Number of ADFs in each chromosome & 2  \\
			
			\bottomrule
		\end{tabular}
	}
\end{table}

\textbf{Deep reinforcement learning methods:}
To verify the effectiveness of our proposed framework, the Double DQN method~\cite{van2016deep} (the detail of the method is in Section~\ref{DDQN}) is implemented as one of the baseline methods in the experiments, where the agent's Q-network is implemented by the neural network with fully-connected layers.

\subsubsection{Training Settings}
The proposed method is trained in an end-to-end way with the embedding dimension $d_h=64$, and the embeddings are learned with 3 massaging and updating iterations. After the feature fusion in multi-head attention training, the embeddings for site nodes are concatenated with 64-dimensional mean pooling of sensor nodes' embeddings. Then the vectors are inputted to a one-layer MLP with hidden dimension 128 and ReLU activation function. For the baseline Double DQN method, the Q-network of the agent is implemented by three fully connected layers with 64 hidden units and ReLU activation functions. Except for the above Q-network settings, the proposed method and RL baseline method share the training settings listed in Table~\ref{DRL parameters}. For the hardware, the training of the neural network is conducted on one NVIDIA GeForce RTX 3080.

\begin{table}[htbp]
	\centering
	\caption{Hyper-parameters for the DRL-based algorithms}\label{DRL parameters}
		\resizebox*{\linewidth}{!}{
		\begin{tabular}{>{\centering\arraybackslash} c | c | c}
			\toprule
			
			\textbf{Parameter name} &\textbf{Symbol}& \textbf{Value}\\
			\hline
			Batch size & $\mathcal{N}$ & 64  \\
			\hline
			Size of replay buffer & $|D|$ & 50k  \\
			\hline
			Training episodes& $\mathcal{M}$ & 100k  \\
			\hline
			Learning optimizer & - & Adam \\
			\hline
			Learning rate & $\alpha$ &1e-4  \\
			\hline
			Maximal exploration probability & $\epsilon_{init}$ &0.99  \\
			\hline
			Minimal exploration probability & $\epsilon_{fin}$ &0.01  \\
			\hline
			Exploration decaying rate per episode & $\epsilon_{decay}$ & 5e-5  \\
			\hline
			Discount factor & $\gamma$ & 0.98  \\
			\bottomrule
		\end{tabular}
	}
\end{table}

\subsection{Experimental Results}\label{Experimental Results}

\subsubsection{Experiments on Stationary WSNs}
According to the setup of experiments in Section~\ref{section5}, we conduct a performance comparison between the proposed method and the baseline methods. In our experiments, the mathematical programming-based method is also used to find a solution with high quality. The mathematical programming-based method is implemented by the GNU Linear Programming Kit package (GLPK-4.65)\footnote{\url{https://www.gnu.org/software/glpk/}}.

\begin{table*}[htbp]
	\centering
	\caption{Comparison On Stationary Wireless Sensor Networks range from map 1 to map 5}\label{static results1}

\resizebox*{\linewidth}{!}{
		\begin{tabular}{l  l | c | c | c | c | c | c |  c | c |  c | c }
			\toprule
		
			\multicolumn{2}{l|}{\multirow{2}{*}{Methods}}& \multicolumn{2}{c|}{Map 1}
			& \multicolumn{2}{c|}{Map 2}& \multicolumn{2}{c|}{Map 3}  & \multicolumn{2}{c|}{Map 4} & \multicolumn{2}{c}{Map 5}\\
			\cline{3-12}   
			& &$\overline{T}$ & $C_t(s)$ & $\overline{T}$ & $C_t(s)$ & $\overline{T}$ & $C_t(s)$ & $\overline{T}$ & $C_t(s)$ & $\overline{T}$ & $C_t(s)$\\
			
			\hline
			Mathematical method& GLPK & 22.15 & 3.42 & \textbf{56.11} & 18.59 & - & - & - & - & - & -\\
			\hline
			\multirow{2}{*}{Heuristic methods} & ACO & 20.11 & 1.058 & 43.42 & 4.447 & 50.39 & 18.374 & 22.64 & 6.63 & 79.62 & 101.66  \\
			\cline{3-12}   
			&GMRE &17.57 &0.002 &37.56 &0.002 &45.22 &0.007 & 17.2 & 0.006 & 58.4 & 0.035\\
			\hline
			Hyper-heuristic methods & SL-GEP & 19.4 & 0.032 & 42.1 & 0.011 & 50.1 & 0.047 & 20.7 & 0.025 & 73.6 & 0.284 \\
			\hline
			\multirow{2}{3.5cm}{Deep Reinforcement\\ learning methods} & DDQN & 20.97 & 0.055 & 43.54 & 0.156 & 49.41 & 0.753 & 21.21 & 0.305 & 72.96 & 2.94 \\
			\cline{3-12} 
			&HGFF &\textbf{32.98} &0.065 &47.27 &0.225 &\textbf{53.34} &0.844 & \textbf{25.15} & 0.418 & \textbf{83.79} & 3.842\\
		
			\bottomrule
		\end{tabular}
	}
\end{table*}
\begin{table*}[htbp]
	\centering
	\caption{Comparison On Stationary Wireless Sensor Networks range from map 6 to map 10}\label{static results2}

	\resizebox*{\linewidth}{!}{
		\begin{tabular}{l  l | c | c | c | c | c | c |  c | c |  c | c }
			\toprule
			
			\multicolumn{2}{l|}{\multirow{2}{*}{Methods}}& \multicolumn{2}{c|}{Map 6}
			& \multicolumn{2}{c|}{Map 7}& \multicolumn{2}{c|}{Map 8}  & \multicolumn{2}{c|}{Map 9} & \multicolumn{2}{c}{Map 10}\\
			\cline{3-12}   
			& &$\overline{T}$ & $C_t(s)$ & $\overline{T}$ & $C_t(s)$ & $\overline{T}$ & $C_t(s)$ & $\overline{T}$ & $C_t(s)$ & $\overline{T}$ & $C_t(s)$\\
			
			\hline
			Mathematical method& GLPK & - & - & - & - & - & - & - & - & - & -\\
			\hline
			\multirow{2}{*}{Heuristic methods} & ACO & 30.22 & 25.713 & 28.67 & 7.669 & 47.32 & 20.147 & 35.13 & 53.293 & 35.37 & 85.36  \\
			\cline{3-12}   
			&GMRE &19.75 &0.025 &18.8 &0.04 &45.13 &0.07 & 17.6 & 0.029 & 22.1 & 0.082\\
			\hline
			Hyper-heuristic methods & SL-GEP & 24.25 & 0.09 & 27 & 0.022 & 46.9& 0.047 & 32 & 0.23 & 34.2 & 0.525 \\
			\hline
			\multirow{2}{3.5cm}{Deep Reinforcement\\ learning methods} & DDQN & 28.05 & 1.3 & 50.95 & 0.733 & 46.9 & 0.72 & 31.54 & 3.52 & 34.58 & 9.61 \\
			\cline{3-12} 
			&HGFF &\textbf{32.2} &1.64 &\textbf{54.48} &0.786 &\textbf{50.98} &0.915 & \textbf{36.74} & 4.48 & \textbf{38.63} & 10.5\\
			
			\bottomrule
			
		\end{tabular}
	}
\end{table*}

The empirical results on ten types of maps with static sensors are shown in Table~\ref{static results1}, \ref{static results2}, with the peak in bold. The value of $\overline{T}$ in each type of map is calculated by the mean of finally achieved lifetimes, which come from ten randomly generated scenarios of each type of map. For HGFF and baseline Double DQN method, we run 20 test episodes and obtain the mean lifetime in each scenario. For the hyper-heuristic method SL-GEP, the lifetime of each scenario is the mean of the lifetimes obtained by the final trained heuristic rule in 20 test episodes. For the random search method ACO, the results are obtained by the mean of 30 searches. The value of $C_t$ is collected from the inference time of the final trained model. In DRL-based methods and hyper-heuristic methods, $C_t$ is collected by the mean of test episodes. Since all the calculation is online in the mathematical method and heuristic methods, $C_t$ records the time from the beginning of the algorithms to the completion of finding its solution.

As shown in the tables, HGFF can obtain the maximal lifetime on all types of maps among the heuristics, hyper-heuristic methods, and DRL-based algorithms. To our surprise, it even outperforms the mathematical method in map 1. It's worth noting that our method is able to get high-quality route for the mobile sink with less time. The stochastic search method ACO performs best between the heuristic methods and hyper-heuristic methods. However, it still cannot obtain the high-quality solutions that HGFF gets, especially since its reasoning takes a long time. In particular, HGFF shows great adaptability through the good results achieved on map 7, which finally achieves twice the network lifetime ratio than the heuristic-based method. Note that the sensors in map 7 are deployed in a rectangle sensing area. The performance comparison in map 7 and map 8 means that HGFF is more suitable in the real world where there are sensing areas of different shapes and numerous occurrences of emergency. Due to the extremely large time consumption and memory usage, the linear programming method implemented by GLPK is only carried on map 1 and map 2, which requires the least computational overhead of all maps.

Overall, our algorithm can consistently outperform all the compared methods on the static WSNs scenarios across all 10 types of maps. However, the MILP methods implemented by GLPK cost too much computational resources and time, which makes it hard to operate in WSNs with more sensors. 

\subsubsection{Experiments on Dynamic WSNs}

The empirical results on ten types of maps with dynamic sensors are shown in Table~\ref{dynamic results1}, \ref{dynamic results2}, with the peak in bold. In dynamic maps, the initial distribution of sensors in each type of map is the same as the stationary map, and the random perturbation method of each sensor follows Section~\ref{sec_5_1_1}. There is an interesting phenomenon that the lifetime obtained by HGFF on most dynamic maps is longer than on corresponding static maps. This means that the location dynamics of sensor nodes have a positive effect on balancing the energy consumption of WSNs. However, heuristics perform worse on some dynamic maps than static ones, which shows that the dynamic map to some extent increases the difficulty of algorithmic decision-making. Accordingly, the inference time of all methods is increased due to the extra computation caused by a more complex simulation of dynamic WSNs. Similar to the performance in stationary maps, HGFF also performs very well in dynamic map 7. 
Besides, with the rising complexity of maps, the reasoning time of most methods gets longer on both static and dynamic maps. In summary, HGFF significantly outperforms the baseline methods on both stationary and dynamic maps.

\begin{table*}[htbp]
	\centering
	\caption{Comparison On Dynamic Wireless Sensor Networks range from map 1 to map 5}\label{dynamic results1} 
\resizebox*{\linewidth}{!}{
		\begin{tabular}{l  l | c | c | c | c | c | c | c | c | c | c }
			\toprule
		
			\multicolumn{2}{l|}{\multirow{2}{*}{Methods}}& \multicolumn{2}{c|}{Map 1}
			& \multicolumn{2}{c|}{Map 2}& \multicolumn{2}{c|}{Map 3} & \multicolumn{2}{c|}{Map 4}& \multicolumn{2}{c}{Map 5}  \\
			\cline{3-12}   
			& &$\overline{T}$ & $C_t(s)$ & $\overline{T}$ & $C_t(s)$ & $\overline{T}$ & $C_t(s)$ & $\overline{T}$ & $C_t(s)$ & $\overline{T}$ & $C_t(s)$ \\
			
			\hline
			\multirow{2}{*}{Heuristic methods} & ACO & 31.89 & 1.03 & 46.97 & 4.43 & 55.62 & 18.24  & 22.58 & 6.66 & 79.73 & 103.73  \\
			\cline{3-12}   
			&GMRE &11.33 &0.003 &10.89 &0.003 &17.88 &0.005  & 8.92 & 0.028 & 19.95 & 0.014\\
			\hline
			Hyper-heuristic methods & SL-GEP & 32.83 & 0.003 & 52.66 & 0.012 & 65.12 & 0.037  & 25.25 & 0.024 & 58.75 & 0.2 \\
			\hline
			\multirow{2}{3.5cm}{Deep Reinforcement\\ learning methods} & DDQN & 32.64 & 0.102 & 50.47 & 0.383 & 56.28 & 1.31  & 21.05 & 0.38 & 83.75 & 5.62 \\
			\cline{3-12} 
			&HGFF &\textbf{35.43} &0.104 &\textbf{54.08} &0.313 &\textbf{65.78} &1.33  & \textbf{25.49} & 0.384 & \textbf{93.47} & 5.875\\
		
			\bottomrule
			
		\end{tabular}
	}
\end{table*}

\begin{table*}[htbp]
	\centering
	\caption{Comparison On Dynamic Wireless Sensor Networks range from map 6 to map 10}\label{dynamic results2}
	\resizebox*{\linewidth}{!}{
		\begin{tabular}{l  l | c | c | c | c | c | c | c | c | c | c }
			\toprule
		
			\multicolumn{2}{l|}{\multirow{2}{*}{Methods}}& \multicolumn{2}{c|}{Map 6}
			& \multicolumn{2}{c|}{Map 7}& \multicolumn{2}{c|}{Map 8} & \multicolumn{2}{c|}{Map 9}& \multicolumn{2}{c}{Map 10}  \\
			\cline{3-12}   
			& &$\overline{T}$ & $C_t(s)$ & $\overline{T}$ & $C_t(s)$ & $\overline{T}$ & $C_t(s)$ & $\overline{T}$ & $C_t(s)$ & $\overline{T}$ & $C_t(s)$ \\
			
			\hline
			\multirow{2}{*}{Heuristic methods} & ACO & 28.48 & 27.09 & 28.79 & 7.69 & 58.98 & 20.28  & 29.72 & 47.38 & 42.5 & 84.72  \\
			\cline{3-12}   
			&GMRE &11.7 &0.016 &19.4 &0.037 &45.8 &0.028  & 17.1 & 0.053 & 20.2 & 0.089\\
			\hline
			Hyper-heuristic methods & SL-GEP & 20 & 0.051 & 32.7 & 0.021 & 66.7 & 0.045  & 20.67 & 0.085 & 38.5 & 0.12 \\
			\hline
			\multirow{2}{3.5cm}{Deep Reinforcement\\ learning methods} & DDQN & 31.47 & 1.93 & 57.83 & 1.24 & 64.55 & 1.21  & 35.22 & 4.84 & 42.74 & 12.27 \\
			\cline{3-12} 
			&HGFF &\textbf{34.86} &2.54 &\textbf{61.14} &1.224 &\textbf{68.16} &1.35  & \textbf{42.94} & 6.56 & \textbf{46.82} & 13.07\\
		
			\bottomrule
			
		\end{tabular}
	}
\end{table*}

\subsubsection{Ablation Study}
In order to probe the efficiency of HGFF, we conduct ablation experiments to investigate the influence of introducing type embeddings into initial features and the necessity of attention-based feature fusion. To explore the performance improvements brought by the two components, three methods are tested on several types of static and dynamic maps: HGFF without feature fusion, HGFF without both type embedding and feature fusion, and HGFF. The experimental results of the three methods on randomly sampled maps are summarized in Table~\ref{ablation results}.

\begin{table*}[htbp] 
	\centering
	\caption{Ablation studies for HGFF}\label{ablation results}
	\resizebox*{\linewidth}{!}{
	\begin{tabular}{c | c | c | c | c | c | c }
		\toprule

		Map & Map 1 & Map 4 & Map 7 & dyn Map 1  &  dyn Map 4 &  dyn Map 7    \\
		\hline
		HGFF & 32.98  & 25.15 & 54.48 & 35.43 & 25.49  & 61.14 \\
		w.o.feature fusion &31.83  &23.55  & 52.94 & 34.52 & 24.1  & 60.16  \\
		w.o.node type embedding + w.o.feature fusion  & 31.71 & 23.43  & 52.6 & 34.29  & 23.82 & 60.16 \\

		\bottomrule
				
	\end{tabular}
}

\end{table*}

\begin{figure}[htbp]
	\centering
	\includegraphics[scale=0.63]{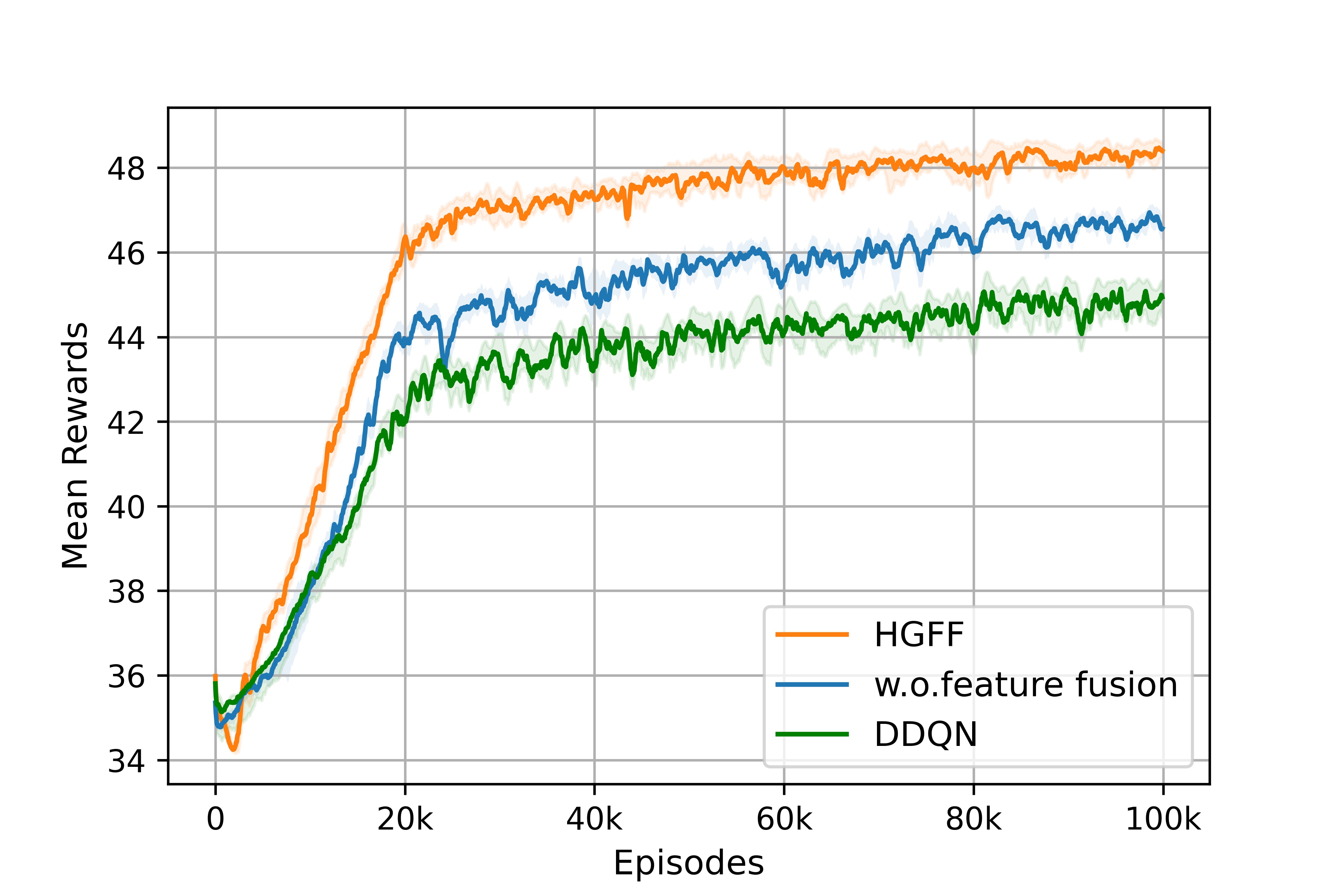}
 \vspace{-3mm}
	\caption{Mean rewards of HGFF, HGFF w.o feature fusion and baseline DDQN method during training on map 1, [25\%,75\%] percentile interval is shaded.}\label{ablation_plot}
\end{figure}

As we can see, the performance gap between the HGFF without the feature fusion method and the HGFF without the two components method shown in the third row and the second row reflects the improvement brought by type embedding. What's more, the comparison between  HGFF without the feature fusion method and HGFF shows the effectiveness of the feature fusion mechanism. The results show that both two components consistently boost performance. The enhancement resulting from feature fusion is evidently larger. Whereas the type embedding only brings slight improvements. The reason for this phenomenon could be that the initial node features already encompass the heterogeneity information.

Take one scenario of map 1 as an example, we visualize the training process of HGFF, HGFF without feature fusion method, and baseline DDQN method. The mean rewards curve of the methods in training is shown in Fig.~\ref{ablation_plot}, and all the experimental results are performed at least 5 runs with different seeds. From the comparison of the baseline DDQN implemented by fully connected networks and HGFF without the feature fusion method, we find that heterogeneous graph learning plays an important role in improving performance. Besides, feature fusion is also very useful to enhance the quality of node representations, which leads a  further improvement.

\begin{figure*}[htbp]
	
	\centering
	\subfigure[The sink movement path obtained by  HGFF and SL-GEP]{
		\begin{minipage}{\linewidth}
			\centering
			\includegraphics[scale=0.35]{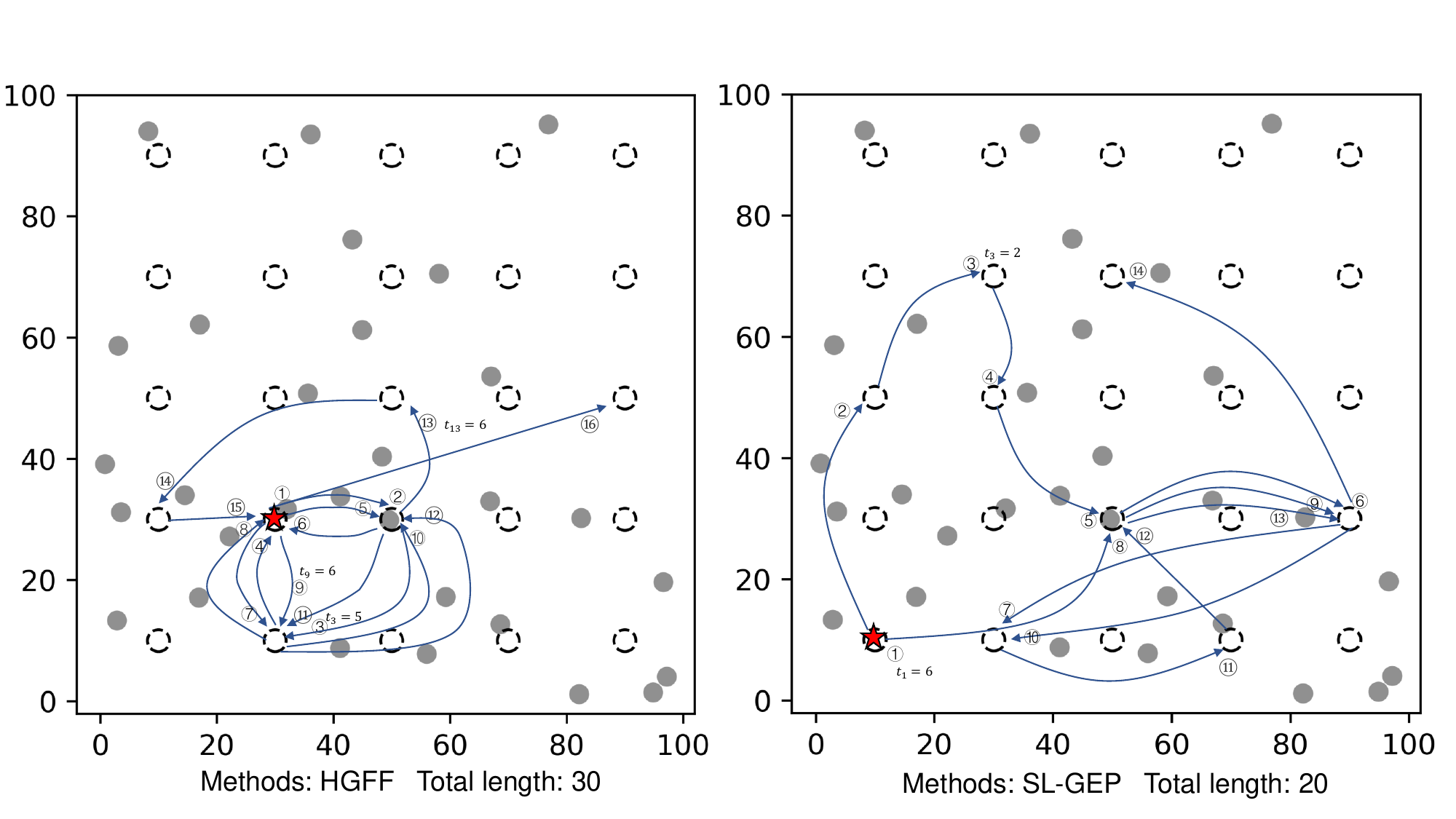}
		\end{minipage}
	}
	\subfigure[The sink movement path obtained by  HGFF and DDQN]{
		\begin{minipage}{\linewidth}
			\centering
			\includegraphics[scale=0.35]{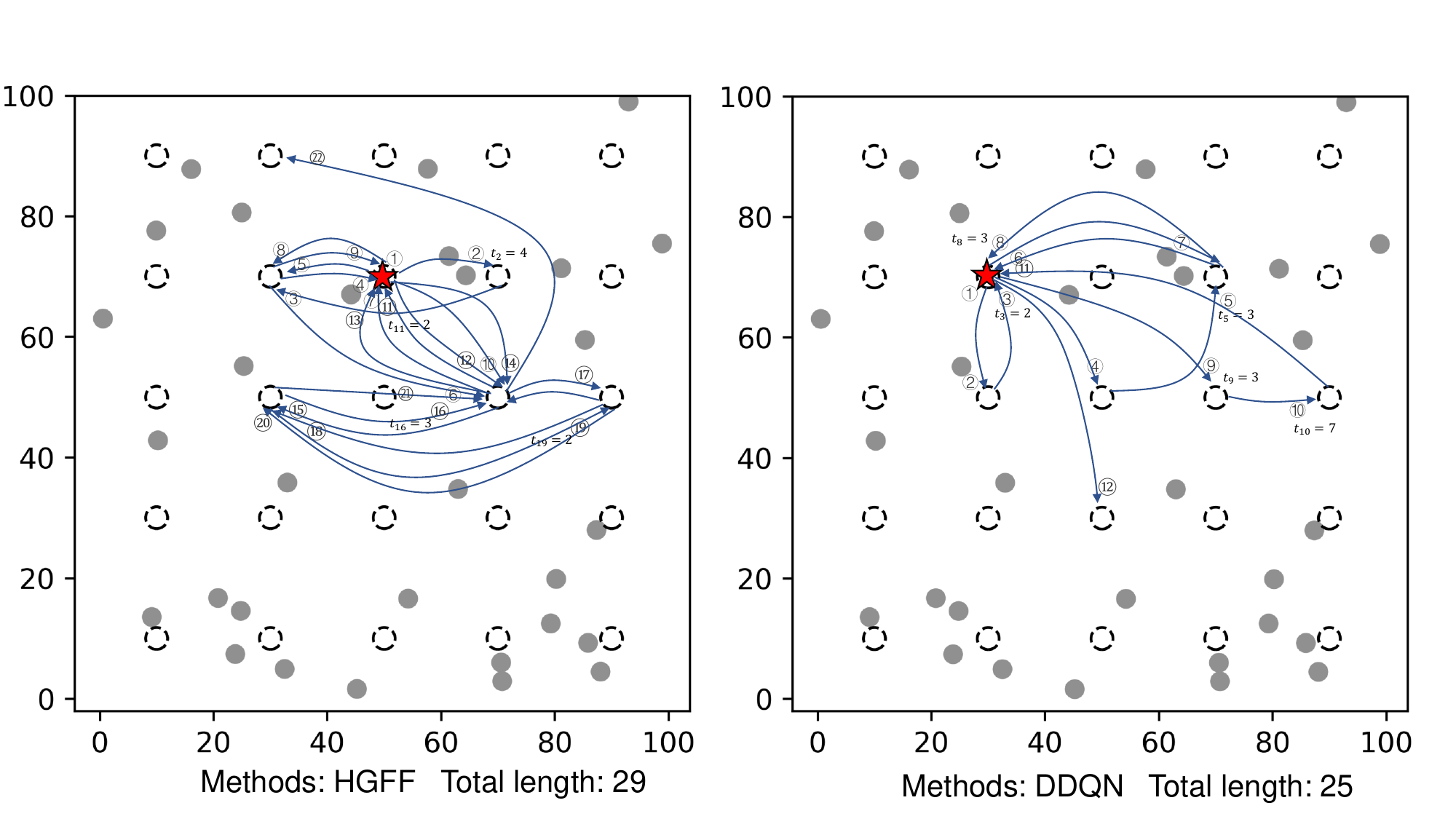}

		\end{minipage}
	}
	\subfigure[The sink movement path obtained by  HGFF and ACO]{
		\begin{minipage}{\linewidth}
			\centering
			\includegraphics[scale=0.35]{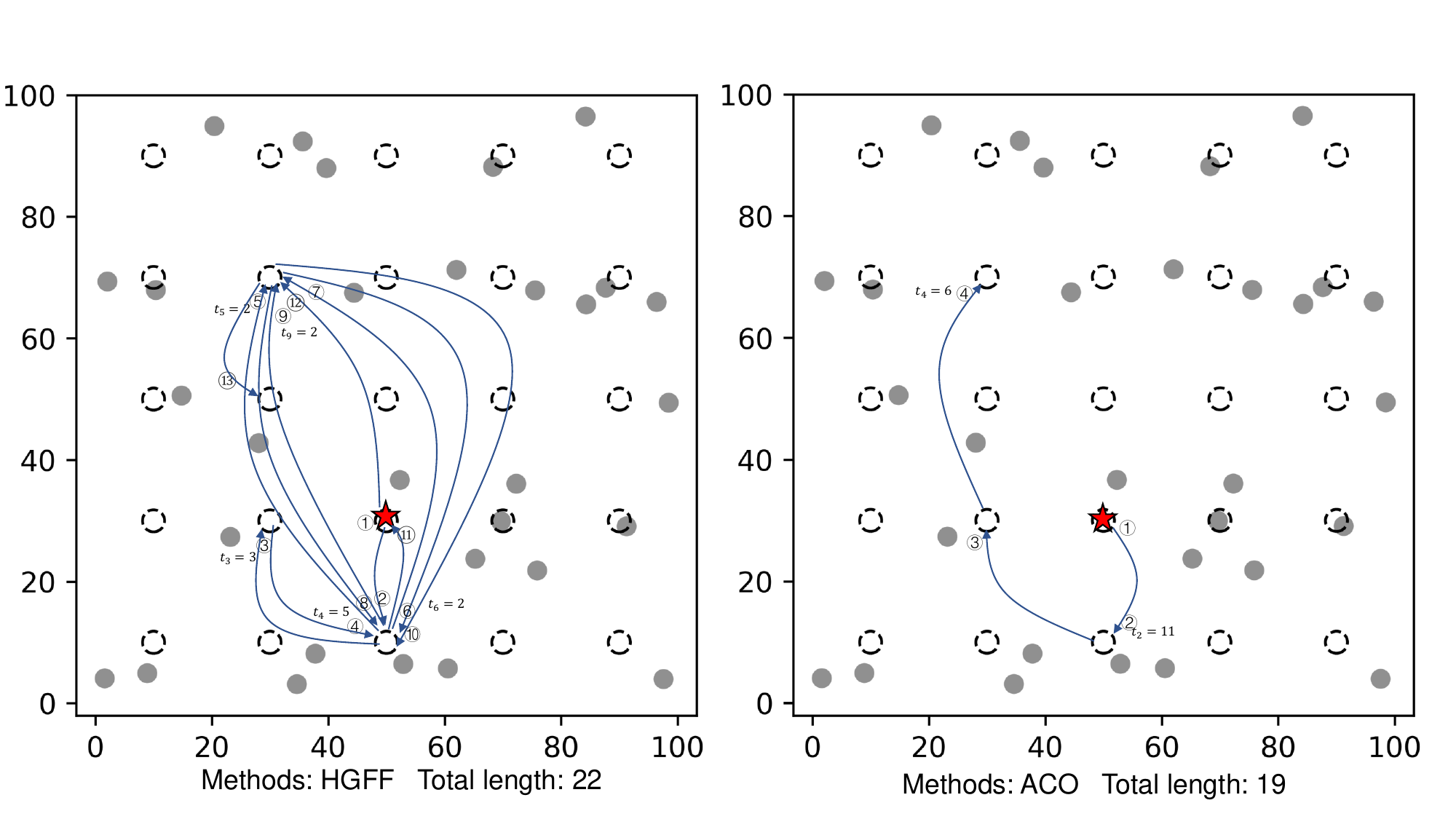}		
		\end{minipage}
	}
	\caption{The visualization of sink movement path obtained by HGFF and baseline methods }
	\label{visualization_routes}
\end{figure*}

\subsubsection{Visualization of the sink movement path}

According to the final trained model, we also visualize the movement path of the sink, which contains the selected site in sequential and the sojourn time on the site. We sampled several scenarios from map 1 which is deployed with the minimum number of sensors. In order to intuitively reflect the final performance of HGFF on maximizing lifespan, the route of sink constructed by baseline algorithms is also plotted for comparison. The movement paths of the sink constructed by HGFF and baseline methods on different scenarios are shown in  Fig.~\ref{visualization_routes}.

The red five-pointed star denotes the sink, the dashed circle represents sites and other gray solid points refer to sensors. The arrow reflects the sink's direction of movement. The number at the end of the arrow represents the movement order of the sink, and $t_i$ indicates the sojourn time to stay at the position according to the movement order $i$. If there is no indication next to the arrow, it means $t_i=1$. It's worth mentioning that before the running of the simulation of WSNs, the sink stays at the site that is closest to the center of the map. As we can see, HGFF can get the maximal route length among all the existing algorithms.

\section{Conclusion}\label{section6}
In this paper, a novel DRL-based framework called HGFF is proposed to address the NP-hard challenge of constructing the optimal route of the mobile sink to maximize the lifetime of WSN. Modeling the problem as an optimization on the heterogeneous graph, we introduce the learnable type embedding in graph learning to emphasize the heterogeneity of nodes. Moreover, the multi-head attention technique is leveraged to further learn the relativity information between the heterogeneous nodes. The extensive experimental results show that our method is capable of learning solutions of good quality in a brief time without human knowledge. Besides, the proposed method also has some reference significance for the problem of deciding the movement of the sink to different cluster heads in hierarchical routing protocol-based WSNs. 

In the future, we'd like to further study the problem of maximizing the lifetime of WSN in which exists more than one sink. In addition,  constructing the optimal data transmission route from sensors to the sink automatically is also an interesting future research direction.

\bibliographystyle{IEEEtran}
\bibliography{refs}

\begin{thebibliography}{10}
\providecommand{\url}[1]{#1}
\csname url@samestyle\endcsname
\providecommand{\newblock}{\relax}
\providecommand{\bibinfo}[2]{#2}
\providecommand{\BIBentrySTDinterwordspacing}{\spaceskip=0pt\relax}
\providecommand{\BIBentryALTinterwordstretchfactor}{4}
\providecommand{\BIBentryALTinterwordspacing}{\spaceskip=\fontdimen2\font plus
\BIBentryALTinterwordstretchfactor\fontdimen3\font minus \fontdimen4\font\relax}
\providecommand{\BIBforeignlanguage}[2]{{%
\expandafter\ifx\csname l@#1\endcsname\relax
\typeout{** WARNING: IEEEtran.bst: No hyphenation pattern has been}%
\typeout{** loaded for the language `#1'. Using the pattern for}%
\typeout{** the default language instead.}%
\else
\language=\csname l@#1\endcsname
\fi
#2}}
\providecommand{\BIBdecl}{\relax}
\BIBdecl

\bibitem{nellore2016survey}
K.~Nellore and G.~P. Hancke, ``A survey on urban traffic management system using wireless sensor networks,'' \emph{Sensors}, vol.~16, no.~2, p. 157, 2016.

\bibitem{mahamuni2021intrusion}
C.~V. Mahamuni and Z.~M. Jalauddin, ``Intrusion monitoring in military surveillance applications using wireless sensor networks (wsns) with deep learning for multiple object detection and tracking,'' in \emph{2021 International Conference on Control, Automation, Power and Signal Processing (CAPS)}.\hskip 1em plus 0.5em minus 0.4em\relax IEEE, 2021, pp. 1--6.

\bibitem{li2020computational}
S.~Li, B.~Zhang, P.~Fei, P.~M. Shakeel, and R.~D.~J. Samuel, ``Computational efficient wearable sensor network health monitoring system for sports athletics using iot,'' \emph{Aggression and Violent Behavior}, p. 101541, 2020.

\bibitem{basagni2006new}
S.~Basagni, A.~Carosi, E.~Melachrinoudis, C.~Petrioli, and Z.~M. Wang, ``A new milp formulation and distributed protocols for wireless sensor networks lifetime maximization,'' in \emph{2006 IEEE International Conference on Communications}, vol.~8.\hskip 1em plus 0.5em minus 0.4em\relax IEEE, 2006, pp. 3517--3524.

\bibitem{agarwal2021survey}
V.~Agarwal, S.~Tapaswi, and P.~Chanak, ``A survey on path planning techniques for mobile sink in iot-enabled wireless sensor networks,'' \emph{Wireless Personal Communications}, vol. 119, pp. 211--238, 2021.

\bibitem{luo2006mobility}
J.~Luo, ``Mobility in wireless networks: friend or foe: network design and control in the age of mobile computing,'' Ph.D. dissertation, Verlag nicht ermittelbar, 2006.

\bibitem{yun2010maximizing}
Y.~Yun and Y.~Xia, ``Maximizing the lifetime of wireless sensor networks with mobile sink in delay-tolerant applications,'' \emph{IEEE Transactions on mobile computing}, vol.~9, no.~9, pp. 1308--1318, 2010.

\bibitem{behdani2012decomposition}
B.~Behdani, Y.~S. Yun, J.~C. Smith, and Y.~Xia, ``Decomposition algorithms for maximizing the lifetime of wireless sensor networks with mobile sinks,'' \emph{Computers \& Operations Research}, vol.~39, no.~5, pp. 1054--1061, 2012.

\bibitem{tashtarian2014maximizing}
F.~Tashtarian, M.~H.~Y. Moghaddam, K.~Sohraby, and S.~Effati, ``On maximizing the lifetime of wireless sensor networks in event-driven applications with mobile sinks,'' \emph{IEEE Transactions on Vehicular Technology}, vol.~64, no.~7, pp. 3177--3189, 2014.

\bibitem{liang2010prolonging}
W.~Liang, J.~Luo, and X.~Xu, ``Prolonging network lifetime via a controlled mobile sink in wireless sensor networks,'' in \emph{2010 IEEE global telecommunications conference GLOBECOM 2010}.\hskip 1em plus 0.5em minus 0.4em\relax IEEE, 2010, pp. 1--6.

\bibitem{wang2018improved}
J.~Wang, J.~Cao, R.~S. Sherratt, and J.~H. Park, ``An improved ant colony optimization-based approach with mobile sink for wireless sensor networks,'' \emph{The Journal of Supercomputing}, vol.~74, no.~12, pp. 6633--6645, 2018.

\bibitem{maheshwari2021energy}
P.~Maheshwari, A.~K. Sharma, and K.~Verma, ``Energy efficient cluster based routing protocol for wsn using butterfly optimization algorithm and ant colony optimization,'' \emph{Ad Hoc Networks}, vol. 110, p. 102317, 2021.

\bibitem{zhong2012ant}
J.-h. Zhong and J.~Zhang, ``Ant colony optimization algorithm for lifetime maximization in wireless sensor network with mobile sink,'' in \emph{Proceedings of the 14th annual conference on Genetic and evolutionary computation}, 2012, pp. 1199--1204.

\bibitem{zhong2019hyper}
J.~Zhong, Z.~Huang, L.~Feng, W.~Du, and Y.~Li, ``A hyper-heuristic framework for lifetime maximization in wireless sensor networks with a mobile sink,'' \emph{IEEE/CAA Journal of Automatica Sinica}, vol.~7, no.~1, pp. 223--236, 2019.

\bibitem{DBLP:conf/iclr/BelloPL0B17}
I.~Bello, H.~Pham, Q.~V. Le, M.~Norouzi, and S.~Bengio, ``Neural combinatorial optimization with reinforcement learning,'' in \emph{5th International Conference on Learning Representations, {ICLR} 2017, Workshop Track Proceedings}, 2017.

\bibitem{DBLP:conf/iclr/KoolHW19}
W.~Kool, H.~van Hoof, and M.~Welling, ``Attention, learn to solve routing problems!'' in \emph{7th International Conference on Learning Representations, {ICLR} 2019}, 2019.

\bibitem{khalil2017learning}
E.~Khalil, H.~Dai, Y.~Zhang, B.~Dilkina, and L.~Song, ``Learning combinatorial optimization algorithms over graphs,'' \emph{Advances in neural information processing systems}, vol.~30, 2017.

\bibitem{barrett2020exploratory}
T.~Barrett, W.~Clements, J.~Foerster, and A.~Lvovsky, ``Exploratory combinatorial optimization with reinforcement learning,'' in \emph{Proceedings of the AAAI Conference on Artificial Intelligence}, vol.~34, 2020, pp. 3243--3250.

\bibitem{bengio2021machine}
Y.~Bengio, A.~Lodi, and A.~Prouvost, ``Machine learning for combinatorial optimization: a methodological tour d’horizon,'' \emph{European Journal of Operational Research}, vol. 290, no.~2, pp. 405--421, 2021.

\bibitem{9471008}
G.~Wu, M.~Fan, J.~Shi, and Y.~Feng, ``Reinforcement learning based truck-and-drone coordinated delivery,'' \emph{IEEE Transactions on Artificial Intelligence}, vol.~4, no.~4, pp. 754--763, 2023.

\bibitem{forster2009clique}
A.~Forster and A.~L. Murphy, ``Clique: Role-free clustering with q-learning for wireless sensor networks,'' in \emph{2009 29th IEEE International Conference on Distributed Computing Systems}.\hskip 1em plus 0.5em minus 0.4em\relax IEEE, 2009, pp. 441--449.

\bibitem{mustapha2017energy}
I.~Mustapha, B.~M. Ali, A.~Sali, M.~F.~A. Rasid, and H.~Mohamad, ``An energy efficient reinforcement learning based cooperative channel sensing for cognitive radio sensor networks,'' \emph{Pervasive and Mobile Computing}, vol.~35, pp. 165--184, 2017.

\bibitem{soni2018novel}
S.~Soni and M.~Shrivastava, ``Novel learning algorithms for efficient mobile sink data collection using reinforcement learning in wireless sensor network,'' \emph{Wireless Communications and Mobile Computing}, vol. 2018, pp. 1--13, 2018.

\bibitem{krishnan2021reinforcement}
M.~Krishnan and Y.~Lim, ``Reinforcement learning-based dynamic routing using mobile sink for data collection in wsns and iot applications,'' \emph{Journal of Network and Computer Applications}, vol. 194, p. 103223, 2021.

\bibitem{watkins1992q}
C.~J. Watkins and P.~Dayan, ``Q-learning,'' \emph{Machine learning}, vol.~8, pp. 279--292, 1992.

\bibitem{hasselt2010double}
H.~Hasselt, ``Double q-learning,'' \emph{Advances in neural information processing systems}, vol.~23, 2010.

\bibitem{khan2013static}
M.~I. Khan, W.~N. Gansterer, and G.~Haring, ``Static vs. mobile sink: The influence of basic parameters on energy efficiency in wireless sensor networks,'' \emph{Computer communications}, vol.~36, no.~9, pp. 965--978, 2013.

\bibitem{ren2015lifetime}
J.~Ren, Y.~Zhang, K.~Zhang, A.~Liu, J.~Chen, and X.~S. Shen, ``Lifetime and energy hole evolution analysis in data-gathering wireless sensor networks,'' \emph{IEEE transactions on industrial informatics}, vol.~12, no.~2, pp. 788--800, 2015.

\bibitem{chang2004maximum}
J.-H. Chang and L.~Tassiulas, ``Maximum lifetime routing in wireless sensor networks,'' \emph{IEEE/ACM Transactions on networking}, vol.~12, no.~4, pp. 609--619, 2004.

\bibitem{van2016deep}
H.~Van~Hasselt, A.~Guez, and D.~Silver, ``Deep reinforcement learning with double q-learning,'' in \emph{Proceedings of the AAAI conference on artificial intelligence}, vol.~30, 2016.

\bibitem{gilmer2017neural}
J.~Gilmer, S.~S. Schoenholz, P.~F. Riley, O.~Vinyals, and G.~E. Dahl, ``Neural message passing for quantum chemistry,'' in \emph{International conference on machine learning}.\hskip 1em plus 0.5em minus 0.4em\relax PMLR, 2017, pp. 1263--1272.

\bibitem{vaswani2017attention}
A.~Vaswani, N.~Shazeer, N.~Parmar, J.~Uszkoreit, L.~Jones, A.~N. Gomez, {\L}.~Kaiser, and I.~Polosukhin, ``Attention is all you need,'' \emph{Advances in neural information processing systems}, vol.~30, 2017.

\bibitem{dorigo2018introduction}
M.~Dorigo and K.~Socha, ``An introduction to ant colony optimization,'' in \emph{Handbook of Approximation Algorithms and Metaheuristics, Second Edition}.\hskip 1em plus 0.5em minus 0.4em\relax Chapman and Hall/CRC, 2018, pp. 395--408.

\end{thebibliography}

\end{document}